\documentstyle[floats,aps]{revtex}
\begin{document}
\draft

\preprint{8-24-93}
\twocolumn[\hsize\textwidth\columnwidth\hsize\csname @twocolumnfalse\endcsname
\title{
Effective Field Theory of Electron Motion\\
in the Presence of Random Magnetic Flux}

\author{Shou-Cheng Zhang}
\address{
IBM Research Division, Almaden Research Center, San Jose, CA 95120
\\ and\\
Department of Physics, Stanford University, Palo Alto, CA 94305}

\author{Daniel P. Arovas}
\address{
Department of Physics, UC at San Diego, La Jolla, CA 92093}

\date{\today}

\maketitle

 \begin{abstract}
  We construct a nonlinear $\sigma$ model to describe a system
  of non-interacting electrons propagating in the presence of
  random magnetic flux.
  We find a term describing the long
  ranged logarithmic interaction between the topological density
  of the non-linear sigma model, and argue that this could give
  rise to a Kosterlitz-Thouless transition from the localized phase to
  a phase with power law correlations and continuously varying
  conductances.  We provide a physical interpretation of our results
  in terms of the scattering of edge states of the magnetic domains
  in different regions.
 \end{abstract}

\vskip2pc]

 \narrowtext
The localization problem in two dimensions has been investigated extensively
in the past twenty years\cite{leeram}.
   Recently there has been considerable interest in studying the
   problem of noninteracting two-dimensional
   electrons propagating in the presence
   of random magnetic flux.  Physically this problem arises in the
   study of the quantum Hall systems near even denominator filling
   fractions\cite{kz,hlr,kwaz} and could also be related to holon
   propagation in chiral spin liquid\cite{chiral}
and the gauge theory describing
   hole motion in an antiferromagnet\cite{naglee}.  Single particle
properties, such as the one-body Green's function and the
density of states, have been addressed\cite{pz,altiof,gws,km}.  A more
difficult question to assess is that of localization.
Na{\"\i}vely, one would assume that the conventional unitary class
sigma model description holds, in which case the
   beta function vanishes to the first order, but is non-vanishing
   and negative to the second order\cite{hikami}, indicating weak
   localization.
For the random flux tight binding model,
\begin{equation}
H=-t_0\sum_{\langle ij\rangle} e^{i a_{ij}}|i\rangle \langle j|+
 e^{-i a_{ij}}|j\rangle \langle i|
\label{eq1}
\end{equation}
with the phase $e^{ia_{ij}}$ on each link uniformly distributed on the
unit circle, numerical calculations
in squares and long strip geometries
   are consistent with the fact that states are localized near
   the bottom of the band\cite{sugnag,ahk,kwaz}.  However, as the
   energy approaches the band center, the localization length
   increases rapidly, and the interpretation of numerical data
   becomes problematic.  In refs.
   \cite{kz,ahk,kwaz} it is was argued that states near the band center
   are extended, whereas ref.\cite{sugnag} argues that all states are
   localized. In view of the rapid increase of the localization length and
   the limit of sample sizes in numerical calculations, it is important
   to develop analytical methods to address the issue of localization.

   In this work, we construct a field theory to describe the
   random flux model.  We envision a system with both random magnetic
flux in addition to a random scalar potential.
Our starting point is the non-linear sigma model of
   the unitary ensemble which is appropriate for the present case with
   broken time reversal invariance.
In the lattice model defined by equation (1), the random flux is correlated
only over the range of an elementary plaquette. However, for the field theory
formulation it is more convenient to consider the opposite limit where the
correlation length of the random magnetic field is large compared to the
elastic mean free path. While these two models addresse two different
limits, we hope that the basic physics continuously interpolates between them.
The basic degrees of freedom of the field theory
   model are the matrices $Q$ defined on the coset space of $U(2m)/U(m)\times
   U(m)$; in the localization problem we are interested in the limit
   of vanishing $m$.  One remarkable feature of this field theory
   is the existence of the topological excitation called instantons.
   It was argued in ref.\cite{llp} that in the presence of a uniform
   magnetic field, there is a term describing the coupling of the Hall
   conductance to the instanton density. This topological term is responsible
   for the existence of extended states necessary for the integer quantum Hall
   effect. In the case of random magnetic flux with no net average, such a
   topological term can not exist. However, it is perfectly sensible to think
   of a ``local topological angle'' associated with a given magnetic domain,
   which is coupled to the local instanton density.
   We are able to demonstrate that there is a finite stiffness to changes
   this local topological angle. This leads to an effective long ranged
   logarithmic interaction between the instanton density. We further argue
   that this long ranged interaction could lead to a Kosterlitz-Thouless
   transition between the free and confined phase of the instantons. In the
   former phase all states are localized whereas extended states exist in the
   later case and there is gapless excitations in the system which gives
   power law correlations.

   Physically, one can think of these gapless excitations as the edge states
   of the fluctuating magnetic domains. These extended edge states can exist
   over a distance larger than the localization length, just as in the case
   of the quantum Hall effect. These edge states are confined in the one
   dimensional network where the magnetic field changes sign. Since the
magnetic
   field is randomly distributed with a zero average, this network always
percolate
  the entire system.
  Of extreme importance is the scattering of
   the edge states with different sense of rotation, or different chirality.
   In the phase where this scattering is relevant, a mass gap opens up and
   all states are localized. In the phase where the scattering is irrelevant,
   the edge states remain gapless and carry currents. There are
   power law correlations
   and a finite, continuously varying conductance.

   We begin with the non-linear sigma model derived by Levine, Libby
  and Pruisken\cite{llp} for the
   case of electrons propagating in a random scalar potential and an uniform
   magnetic field. The Lagrangian takes the following form:
  \begin{equation}
  {\cal L}  =  \frac{1}{8} \sigma_{xx}\mathop{\rm Tr}\,
  (\partial_{\mu} Q)^2 -\frac{1}{8}
  \sigma_{xy} \mathop{\rm Tr}\,
  \epsilon^{\mu\nu} Q\, \partial_\mu Q \,\partial_\nu Q
  \label{eq2}
  \end{equation}
 where $Q$ is a matrix defined on the coset space of $U(2m)/U(m)\times U(m)$,
 and $\sigma_{xx}$ and $\sigma_{xy}$ are the longitudinal and the Hall
 conductances respectively, defined as
  \begin{eqnarray}
  \sigma_{xx} & = & \langle\hat\sigma_{xx}\rangle = - \langle\mathop{\rm Tr}\,
  \Pi_x (G^+ - G^-) \Pi_x
  (G^+ - G^-)\rangle_{\rm imp} \nonumber \\
  \sigma_{xy} & = & \sigma_{xy}^{\rm I} + \sigma_{xy}^{\rm II} \nonumber \\
  \sigma_{xy}^{\rm I} & = & \langle\hat\sigma_{xy}^{\rm I}\rangle
  = - \langle\mathop{\rm Tr}\,
  \epsilon_{\mu\nu} \Pi_\mu G^+ \Pi_\nu
  G^-)\rangle_{\rm imp} \nonumber \\
  \sigma_{xy}^{\rm II} & = & \langle\hat\sigma_{xy}^{\rm II}\rangle =
  i \langle\mathop{\rm Tr}\, \epsilon_{\mu\nu} r_\mu
  \Pi_\nu (G^+ - G^-)\rangle_{\rm imp}
\label{eq3}
  \end{eqnarray}
  In the above formula, $r_\mu$ denotes the position operator and
  $\Pi_\mu$ denotes the gauge invariant momentum operator.
  $G^+$ and $G^-$ are the advance and retarded one particle Green's
  functions respectively.
  $\langle\ldots\rangle_{\rm imp}$ denotes impurity average
  whereas $\langle\ldots\rangle_{\rm imp}$ denotes
  both quantum mechanical trace and impurity average. The second term
  in the Lagrangian is called a topological term since for any
  smooth field configurations,
  \begin{equation}
  \int\!\!d^2 r\, \mathop{\rm Tr}\,
\epsilon^{\mu\nu}\, Q \partial_\mu Q\, \partial_\nu Q
  = 16 \pi i\, n
\label{eq4}
  \end{equation}
  where $n$ is an integer, called the Pontrjagin index. Different
  $n$ label different topological sectors with a given net instanton
  number. The presence of the topological term violates both time
  reversal and parity, and gives a handness to the two dimensional
  plane. Such a term is not allowed on symmetry grounds for the
  random flux problem with no net averaged magnetic field.

$\sigma_{xy}^{\rm II}$ is a one body operator that measures the magnetization
at the fermi energy, it receives only contribution from the edge states
\cite{llp,streda}. In the random field problem, its average vanishes,
but its correlation due to the fluctuating magnetic domains is long
ranged. To see this, we notice that its contribution to
  the topological term arises in a cumulant expansion of
  \begin{eqnarray}
  \langle\exp\frac{1}{8}\int\!\!d^2\, r\hat\sigma_{xy}^{\rm II}(r)\rho(r)
  \rangle =
  1 + \frac{1}{8}\int\!\! d^2 r \langle\hat\sigma_{xy}^{\rm II}(r)\rangle
  \rho(r) +
  \nonumber \\
  \frac{1}{128}\int\!\!d^2 r\, d^2 r'\, \langle\hat\sigma_{xy}^{\rm II}(r)
  \hat\sigma_{xy}^{\rm II}(r')\rangle\,
  \rho(r)\,\rho(r') + \ldots
  \label{eq5}
  \end{eqnarray}
  where $\rho(r) =
 \mathop{\rm Tr}\,
\epsilon^{\mu\nu}\, Q \,\partial_\mu Q \,\partial_\nu Q$
 is the local Pontrjagin density.
  In the presence of an uniform magnetic field, the second term
  in nonzero, reexponentialting it gives the topological term
  in the Lagrangian of equation (\ref{eq2}).
  In this case, one has to
  reorganize the cumulant expansion into an expansion in
  $\hat\sigma_{xy}^{\rm II} - \langle\hat\sigma_{xy}^{\rm II}\rangle$.
  It can be easily shown that the
  correlation function of this operator is short ranged, therefore
  keeping higher order terms in the cumulant expansion would only
  give rise to irrelevant higher derivative terms. In the case of
  random flux with no net average, the second term in equation (\ref{eq5})
  vanish identically. One is left to calculate the two point
  correlation function of local $\hat\sigma$ operators. This correlation
  function can be calculated using a Ward identity relating it to
  the current current correlation function. Define the local current
  operator to be $J_\mu (r) = i \Pi_\mu (r)(G^+ - G^-)$. For the purpose
  of calculating DC transport properties one can assume that the
  current is transverse, {\it i.e\/} $\partial_\mu J_\mu = 0$.
  Furthermore let us assume that the current operator can be expanded
  in terms of its moments. Under these two assumptions,
  \begin{equation}
  J_\mu (q) = -\frac{i}{2}\epsilon_{\mu\nu}\, q_\nu
  \,\hat\sigma_{xy}^{\rm II}(q)
  + o(q^2)
  \label{eq6}
  \end{equation}
  This is nothing but the familiar identity $\vec J = {1\over 2} \vec
  \nabla \times \vec M$ in electromagnetism, expressing current
  in terms of the magnetic moment, which in our case is just
  $\sigma_{xy}^{\rm II}$. Since
  this is an operator identity, it can be used to derive
  the following identity between the correlation functions:
  \begin{equation}
  \langle J_\mu (q) J_\nu(-q)\rangle = \frac{1}{4}(\delta_{\mu\nu}\,
  q^2 - q_\mu\, q_\nu) \langle\hat\sigma_{xy}^{\rm II}(q)\,
  \hat\sigma_{xy}^{\rm II}(-q)\rangle + o(q^4)
  \label{eq7}
  \end{equation}
  In the limit of $q\to 0$, the transverse current current correlation
  of the left hand side approaches $\sigma_{xx}$, therefore we
  obtain
  \begin{equation}
  \langle\hat\sigma_{xy}^{\rm II}(q)\, \hat\sigma_{xy}^{\rm II}(-q)\rangle
   = {4 \sigma_{xx} \over q^2}
  \label{eq8}
  \end{equation}
  Since $\sigma_{xy}^{\rm I}$ is already a two body operator, its correlation
  function enters higher order cumulant expansion which will not be
  considered here.
  Inserting (\ref{eq8}) into equation (\ref{eq5})
  and reexponentialte the
  third term, one obtains the following expression for the field
  theory action of the random flux model:
  \begin{equation}
  {\cal S}  =  \frac{\sigma_{xx}}{8}\int\!\! d^2 r\
  \mathop{\rm Tr}\,(\partial_{\mu} Q)^2 -
  \frac{\sigma_{xx}}{32} \int\!\! {d^2 q\over (2\pi)^2}\, q^{-2}\,
  \rho(q)\,\rho(-q)
  \label{eq9}
  \end{equation}
  Thus we see that while there is no topological term in the action
  for the random flux problem, there is a term describing the long
  ranged interaction of the topological density.

  The physical meaning of this nonlocal interaction term becomes
  transparent if one introduces an additional boson field to
  decouple the nonlocal term. The action becomes local after this
  decoupling:
\begin{equation}
  {\cal L}  =  \frac{\sigma_{xx}}{8} \mathop{\rm Tr}\,
(\partial_{\mu} Q)^2 +
  {8\over \sigma_{xx}}\,
  (\partial_\mu \phi)^2 + \phi(r)\, \rho(r)
  \label{eq10}
  \end{equation}
  The above equation is the central result of this paper, we propose
  this action to be the effective field theory model describing the
  random flux problem.
  The last term in the above equation is reminiscent of the topological
  term, except that the constant topological angle
  is now a space dependent
  scalar field, and there is a gradient term for the scalar field.
  The gradient term keeps $\phi$ locally constant. The locally constant
  $\phi$ field can be viewed as the topological angle in a given
  magnetic domain in the random flux problem. Another physical
  interpretation can be obtained when we use the fermionic
  representation of the $\phi$ field. According to the bosonization
  rules, one scalar field can be represented by a Dirac fermi field,
  with the gradient term $(\partial_\mu \phi(r))^2$ replaced by the
  Dirac action $\bar\psi \gamma^\mu  \partial_\mu \psi$. The
  equation of motion for the bose field $\phi$ is
\begin{equation}
  \partial_\mu \left({16\over \sigma_{xx}}\, \partial_\mu \phi\right) = \rho
  \label{eq11}
  \end{equation}
  The fermionized version of this equation is
  \begin{equation}
  \partial_\mu \left({16 \sqrt{\pi} \over \sigma_{xx}}\,
  \bar\psi\, \gamma_5\, \gamma^\mu\, \psi\right) = \rho
  \label{eq12}
  \end{equation}
  This is the famous chiral anomaly
  equation that arises in the resolution of the $U(1)$ problem
  in the quantum chromodynamics\cite{coleman}. It is also similar to a model
  of holes moving in antiferromagnet background in 1+1 dimensions
  \cite{shankar}. The spin degrees of freedom is described by
  the $O(3)$ nonlinear sigma model whereas the massless fermions
  describes the holes in the antiferromagnetic background.
  These two degrees of freedom are coupled exactly through the
  anomaly equation (\ref{eq12}).
  At the classical level, the two
  chiral components of the Dirac fermion are decoupled since the
  mass term is absent. However, this symmetry is broken at the
  quantum level by the presence of instantons. In the topologically
  trivial sector, the integrated Pontrjagin density vanishes and the
  chiral charge is conserved. The presence of instantons violates the
  conservation of the chiral charge.

  Physically, one can view the two chiral components
  of the Dirac fermion as the gapless edge states of the fluctuating
  magnetic domains\cite{edge}.
  The right chiral fermion can be associated with
  the magnetic domain in the $+ z$ direction, and the left chiral
  fermion can be associated with the magnetic domain in the
  opposite direction. One can visualize the world line of the two
  chiral fermions as the current loops of the edge states.
  The edge states can be extended over a
  distance far larger than the localization length, thus they
  appear to be gapless at the classical level. However, there is
  tunneling between the two chiral components, where the tunneling
  event in our formulation is associated with the instanton, and
  the tunneling rate given by the instanton action. A crucial
  question is whether the chiral anomaly (\ref{eq11}) and (\ref{eq12})
  gives an effective
  mass for the Dirac fermion or if the fermions remain massless even
  in the presence of instantons. Formulated in another way, the
  question is whether the scattering between the two different types
  of edge states is relevant or irrelevant.

  One way to answer this question is to integrate out the $Q$ fields
  in the nonlinear sigma model (\ref{eq10}) and obtain an effective
  theory for the boson field $\phi$. One notices that there is
  a symmetry in the problem: if one shifts the $\phi$ field by a
  constant $\theta$, the gradient term remain unchanged. One picks
  up a contribution
\begin{equation}
  \theta \int\!\! d^2 r\, \rho(r) = 16 \pi i \theta\, n
  \label{eq13}
  \end{equation}
  from the coupling of the $\phi$ field to the topological density.
  The above equality holds because of the quantization of the
  topological charge (\ref{eq4}). For $\theta = {1\over 8}$, this shift
  contributes
  a factor of $e^{i 2\pi n}=1$ to the action, thus $\phi(r)\to
  \phi(r) + {1\over 8}$ is an exact symmetry of the problem that must be
  respected by the effective Hamiltonian in $\phi$. A mass term for
  the $\phi$ field, $m^2 \phi^2$ for example, is forbidden by this symmetry.
  Because of this
  symmetry, the effective action for $\phi$ can be expanded in
  a Fourier series,
\begin{equation}
  {\cal L}_{\rm eff} = {8\over \sigma_{xx}}\, (\partial_\mu \phi)^2
  - 2 y\,\cos (16\pi\phi) +\ldots
  \label{eq14}
  \end{equation}
  The higher harmonics term have higher scaling dimensions compared
  to the lowest harmonic term, and they can be neglected in the
  analysis of critical properties. Thus, general symmetry
  considerations severely restrict the low energy effective action
  for the $\phi$ field to be the sine-Gordon model.
  At this level, $y$ is
  an unknown parameter. While this effective
  action for the $\phi$ field is obtained from general symmetry
  considerations, it can also be derived explicitly from a dilute
  instanton gas approximation.
  It is known that the nonlinear sigma model for the $Q$ matrices
  supports instanton solutions that survives the replica limit
  $m\to 0$\cite{llp}. They make a contribution of
  $y = D \exp(-4\pi \sigma_{xx})$ to the action, where the exponent
  is the classical action of an instanton and $D$ is the fluctuation
  determinant around the classical instanton. Assuming that these
  instantons can be treated as independent variables, one obtains
  the following for the effective action of the $\phi$ field
\begin{eqnarray}
  e^{-S_{\rm eff}[\phi]} & = & \exp\left[ \frac{1}{2}
  \left({16\over \sigma_{xx}}\right) \int\!\! d^2 x\,
  (\partial_\mu \phi)^2 \right.
  \nonumber \\
  && \left.+\sum_{n_+,n_-} {y^{n_+}\over n_+!}\, {y^{n_-}\over n_-!}
  \int\!\! dx_1\cdots dx_{n_+} e^{16\pi i\sum\phi(x_i)}\right.
  \nonumber \\
  &&\left.\times
  \int\!\! d\bar x_1\cdots d\bar x_{n_-}\, e^{-16\pi i\sum\phi(\bar x_i)}
  \right]
  \label{eq15}
  \end{eqnarray}
  the sums and integrals can be done explicitly to give the effective
  action (\ref{eq14}).

  The effective action (\ref{eq14}) is the sine-Gordon model for
  scalar field, and has a well known Kosterlitz-Thouless phase
  transition. After a rescaling of the field $\phi$ one obtains
\begin{equation}
  {\cal L}_{\rm eff} = \frac{1}{2}(\partial_\mu \phi)^2 -
  2 y\ \cos (4\pi \sqrt{\sigma_{xx}} \phi)
  \label{eq16}
  \end{equation}
  this is identical to the effective Lagrangian for the $XY$ model
  \cite{kogut}
  provided one makes the identification of $\sigma_{xx}$ with
  the inverse temperature $\beta = 1/kT$ and $y$ with the fugacity
  of the $XY$ vortices. The Kosterlitz-Thouless transition occurs
  at $\sigma_{xx}^c = 2/\pi$. The conductance $\sigma_{xx}$ which
  enters the effective field theory is the ``bare'' conductance
  calculated within the self-consistent Born approximation.
  $\sigma_{xx}^c$ divides the regime where the cosine term is
  relevant from the one in which the cosine term is irrelevant.
  For $\sigma_{xx} < \sigma_{xx}^c$, one is in the
  high temperature phase where instantons form a free plasma and
  the system is disordered with a finite mass gap. However, for
  $\sigma_{xx} > \sigma_{xx}^c$,
  one is in the low temperature phase in which
  the instantons form tightly bound pairs. There is line of fixed
  points with power law correlations. The massive phase is
  identified with the localized phase in which the scattering between
  the two different chiral edge states are relevant, whereas the
  massless phase is identified with the phase where extended states
  exists and the scattering between the edge states is irrelevant.
  The current is carried by these gapless edge states and they are
  responsible for the finite continuously varying conductance in this
  regime. This picture is perfectly consistent with the recent
  numerical results of ref.\cite{kz,ahk,kwaz}, where it was suggested
that a line of fixed points might exist in a finite region around
the center of the band. The
  conductance was found to be finite and continuously varying in this energy
  range.

However, a word of caution concerning the dilute instanton gas
approximation is in order here. In a model where the massless
fermions are coupled to the gauge fields of the abelian Higgs model
\cite{cdg}, the dilute instanton gas approximation is valid because
the instanton has a well defined size. Consequently
the existence of the Kosterlitz-Thouless phase transition is on a
firm ground. For the case of massless fermions coupled to nonlinear
sigma model, the instantons have arbitrary size and therefore they
can overlap with each other which makes dilute instanton approximation
hard to justify. This question was studied extensively in the context
of the $CP^{n-1}$ model\cite{witten,affleck1,affleck2}. In the large $n$
limit, Witten\cite{witten} has shown that the instanton disappear after
quantum fluctuations are taken into account, and the $\phi$ field is always
massive. However, it is now well established that this large $n$ result
does not hold for $n\leq 2$. For the case of $n=2$, the $CP^{n-1}$ model
is equivalent to the $O(3)$ nonlinear sigma model. From the analogy with
quantum spins chains, it is known that this model has a {\it second order}
phase transition when the topological angle is equal to $\pi$, whereas
the large $n$ theory predicts a {\it first order} phase
transition\cite{affleck1}.
For the case of $n=1$, the $CP^{n-1}$ model is equivalent to the abelian Higgs
model, where the instanton calculus is certainly correct.
Based on this general arguments, Affleck\cite{affleck2} has argued that the
large $n$ result does not hold for $n\leq 2$. It is thus plausible that the
qualitative feature obtained by the instanton calculus is valid in this regime,
and in particular, in the replica limit of $n\rightarrow 0$ that we are
interested
here.


In conclusion we have constructed an effective field theory of the
localization problem in random magnetic flux. This field theory has the
same degrees of freedom as the nonlinear sigma model of unitary
ensemble, but there is a long ranged logarithmic interaction of
the topological density. This model can be reformulated in terms of
a scalar field coupling to the topological density or equivalently,
a Dirac fermi field obeying a chiral anomaly equation. These
additional degrees of freedom can be physically interpreted as the edge
states of the different magnetic domains. Our field theory model
describes the tunneling between the edge states of different chirality
in terms of the instantons of the nonlinear sigma model. The resulting
effective field theory for the scalar field is shown to be a sine Gordon
model from general symmetry considerations and we argue that such a
model has a Kosterlitz-Thouless phase transition from localized states
to extended states with power law correlations. This picture is
consistent with the recent numerical calculations of the random flux
model in which a divergent localization length was found and a finite
region of continuously varying conductance was identified. Our results
have important consequences for the localizations physics by demonstrating
the existence of a finite region of extended states in two dimensions, and
deepen our understanding of the ``Hall metal'' phase around the even
denominator filling fractions in the quantum Hall systems.

The authors would like to acknowledge interesting and stimulating
discussions with J. Chalker, D.H. Lee, H. Levine, S. Libby, N. Read, R.
Shankar and A. Tikofsky.
Part of this work was carried out at the Aspen center for
physics during the workshop on the quantum Hall effect.

   \end{document}